\documentclass[%
superscriptaddress,
showpacs,
amsmath,amssymb,
twocolumn,
prc
floatfix, ]%
{revtex4-1}
\usepackage{color}
\usepackage{hyperref}
\definecolor{mygray}{gray}{.9}
\definecolor{mypink}{rgb}{.99,.91,.95}
\definecolor{mycyan}{cmyk}{.3,0,0,0}
\usepackage{graphicx}
\usepackage{dcolumn}
\usepackage{bm}
\usepackage{amsmath,amssymb}
\usepackage{booktabs}
\usepackage{cases}
\usepackage{multirow}
\usepackage{longtable}

\begin{document}

\title{Study on nuclear $\alpha$-decay energy by an artificial neural network with pairing and shell effects}

 \author{Hong-Qiang You}%
 \affiliation{College of Science, Nanjing University of Aeronautics and Astronautics, Nanjing 210016, China}
 \affiliation{College of Materials Science and Technology, Nanjing University of Aeronautics and Astronautics, Nanjing 210016, China}

 \author{Zheng-Zhe Qu}%
 \affiliation{College of Materials Science and Technology, Nanjing University of Aeronautics and Astronautics, Nanjing 210016, China}
 
 \author{Ren-Hang Wu}%
 \affiliation{College of Materials Science and Technology, Nanjing University of Aeronautics and Astronautics, Nanjing 210016, China}
 
\author{Hao-Ze Su}%
 \affiliation{College of Materials Science and Technology, Nanjing University of Aeronautics and Astronautics, Nanjing 210016, China}

 \author{Xiao-Tao He}%
\email{hext@nuaa.edu.cn}
 \affiliation{College of Materials Science and Technology, Nanjing University of Aeronautics and Astronautics, Nanjing 210016, China}

\date{\today}

\begin{abstract}
We build and train the artificial neural network model (ANN) based on the experimental $\alpha$-decay energy ($Q_{\alpha}$) data. Besides decays between the ground states of parent and daughter nuclei, decays from the ground state of parent nuclei to the excited state of daughter nuclei are also included. By this way, the number of samples are increased dramatically. The results calculated by ANN model reproduce the experimental data with a good accuracy. The root-mean-square (rms) relative to the experiment data is 0.105 MeV. The influence of different input is investigated. It is found that either the shell effect or the pairing effect results in an obvious improvement of the predictive power of ANN model, and the shell effect plays a more important role. The optimal result can be obtained as both the shell and pairing effects are considered simultaneously. Application of ANN model in prediction of the $\alpha$-decay energy shows the neutron magic number at $N=184$, and a possible sub-shell gap around $N=174$ or 176 in the superheavy nuclei region. 
\end{abstract}


\maketitle

\newpage

\section{Introduction}
$\alpha$-decay is one of the most important decay mode of the heavy nuclei. It plays a crucial role in the identification of the newly synthesized superheavy element and provides reliable informations on the nuclear structures. Moreover, the $\alpha$-decay half-life is one of the decisive factors for the stability of the superheavy nuclei~\cite{RevModPhys}. The theoretical $\alpha$-decay half-life is very sensitive to $\alpha$-decay energy which itself can reveal many nuclear structural properties~\citep{PhysRevC.36.456,2003Cluster,PhysRevC.73.061301}. 
 
The $\alpha$-decay energy can be obtained from nuclear mass difference of the involved nuclei, where nuclear mass is calculated by various theoretical mass models \citep{PhysRevC.52.R23,2001Simple,1936Nuclear,PhysRevLett.108.052501,2014Surface,PhysRevLett.102.152503,PhysRevC.93.034337,1995The,2005Relativistic,PhysRevC.73.037303,PhysRevLett.101.122502,PhysRevC.95.044301}. The accuracy of these mass models range from about $3$ MeV for the Bethe-Weizsäcker (BW) model \citep{2008Mutual} to about $0.3$ MeV for the Weizsäcker-Skyrme (WS) model \citep{2014Surface}. For heavy nuclei, an uncertainty of $1$ MeV $Q_{\alpha}$ would lead to $10^{3-5}$ times uncertainty of $\alpha$-decay half life.~\citep{P1997NUCLEAR}. Such accuracy can not meet the request of the $\alpha$-decay half-life investigation, especially for the unknown nuclear mass region, such as superheavy and neutron-rich region. Therefore, accurate description of the known nuclear $\alpha$-decay energy and reliable prediction of the unknown are required indisputably.  

The empirical formula is also applied to investigate $\alpha$-decay energy. The systematics of $\alpha$-decay energy is analyzed within the valence correlation scheme~\cite{JiaJ2021_PRC103_24314}. A formula based on the method of macroscopic model plus shell corrections is proposed to study the $\alpha$-decay energy for superheavy nuclei~\citep{PhysRevC.82.034320,2012Binding}. Based on prediction of nuclear mass, the $\alpha$-decay energy is investigated for transuranium~\citep{PhysRevC.85.054303}. A formula for the relationship between the $\alpha$-decay energy ($Q_{\alpha}$) of superheavy nuclei (SHN) is presented, which is composed of the effects of Coulomb energy and symmetry energy \citep{2011Correlation,PhysRevC.103.024314}. 

In recent years, machine learning has been used in the research of nuclear physics \citep{hao2016Deep,2018Nuclear,2018Bayesian,2019Neutron,E2021Nuclear,SaxenaG2021_JPGNPP48_55103,2019Alpha}. The artificial neural networks is employed on experimental nuclear charge radii and ground-state energies of nuclei~\citep{Akkoyun2013An,hao2016Deep}. In Refs. \citep{2018Nuclear,2018Bayesian,2019Neutron,E2021Nuclear}, the Bayesian neural network (BNN) approach is used to improve the nuclear mass predictions of various models and two-neutron separation energies $S_{2n}$. It constructs a sufficiently complex neural network that can accelerate the calculation of relevant physical quantities with many parameters. 

For the $\alpha$-decay energy, to the best of our knowledge, except for the studies in Refs.~\citep{SaxenaG2021_JPGNPP48_55103} and  \citep{2019Alpha}, it is rarely investigated by using the machine learning method. In Ref. \citep{SaxenaG2021_JPGNPP48_55103}, $\alpha$-decay energy is calculated for the nuclei within the range $82 \leqslant Z \leqslant 118$ by four different machine learning modes which includes XGBoost, Random Forest (RF), Decision Trees (DTs), and Multilayer Perception (MLP) neural network. It is found that XGBoost reproduce better of the experimental $Q_{\alpha}$ values and the root-mean-square deviation is 0.31. In Ref. \citep{2019Alpha}, the BNN model is used to calculate $\alpha$-decay energy for the nuclei within the range $82 \leqslant Z \leqslant 118$. By this way, the $Q_{\alpha}$ value prediction of the Duflo-Zuker mass model has been improved and the root mean square deviation improvement is 72$\%$. In both of the works, only $\alpha$ decays from the ground state of the parent nucleus to the ground state of daughter nucleus are considered. 

In this work, based on the experimental $\alpha$-decay energy data~\citep{timmurphy.org}, we construct an artificial neural network to study the $Q_{\alpha}$ value. In addition to decays which is only involved the ground states of the parent and daughter nuclei, the decays of the ground state of the parent nucleus to the excited state of the daughter nucleus are investigated as well. By this way, the number of samples is increased substantially. We choose four inputs, i.e. mass number (A), proton number (Z), shell effect $P$ and pairing term $\delta$. The accuracy of the description of the experimental $Q_{\alpha}$ is improved. Then the trained ANN model is used to predict the $Q_{\alpha}$ for the superheavy nuclei, and the shell gap at $N=184$ is presented. 

\section{Theoretical Framework}
\label{sec:TheorFrame}
We apply ANN model to calculate the $\alpha$-decay energy. Our model takes the nuclear properties, which have important effect on the $\alpha$-decay energy of nuclei, as input and the $\alpha$-decay energy $Q_{\alpha}$ as the output. Such problems, where the output is a numerical value, are known as the regression problems. Regression is a supervised learning problem where there is an input-x, an output-y, and the task is to learn the mapping from the input to the output \citep{Pnar2013Introduction}. A model in machine learning is assumed as below,

\begin{align}
y_{\omega,b}(\textbf{x})=\phi(\textbf{xw})
\end{align}
where $\phi$(.) is the activation function and $\pmb{\omega}=\left\lbrace {\omega_{i}}\right\rbrace $ are weight parameters. In our study, $y$ corresponds to the output representing prediction for alpha decay energy, and $\pmb{x}=\left\lbrace {x_{i}}\right\rbrace$ is the input data. In the context of machine learning, the parameters $\omega_{i}$ are optimized by minimizing a loss function. Thus, the predictions are obtained as close as possible to the correct values given in the input data.

Multilayer perception (MLP), which is a class of feedforward artificial neural network, was selected as the model to solve the regression problems. During the training phase of MLP, the back propagation algorithm is used for calculation of gradient \citep{1986Learning}. In recent years, a new algorithm called adaptive gradient method is the preferred optimization method. In this study, we used $Adam$ algorithm to train ANN model \citep{2015Deep,Deep}.

\begin{figure*}[htbp]
\includegraphics[scale=0.5]{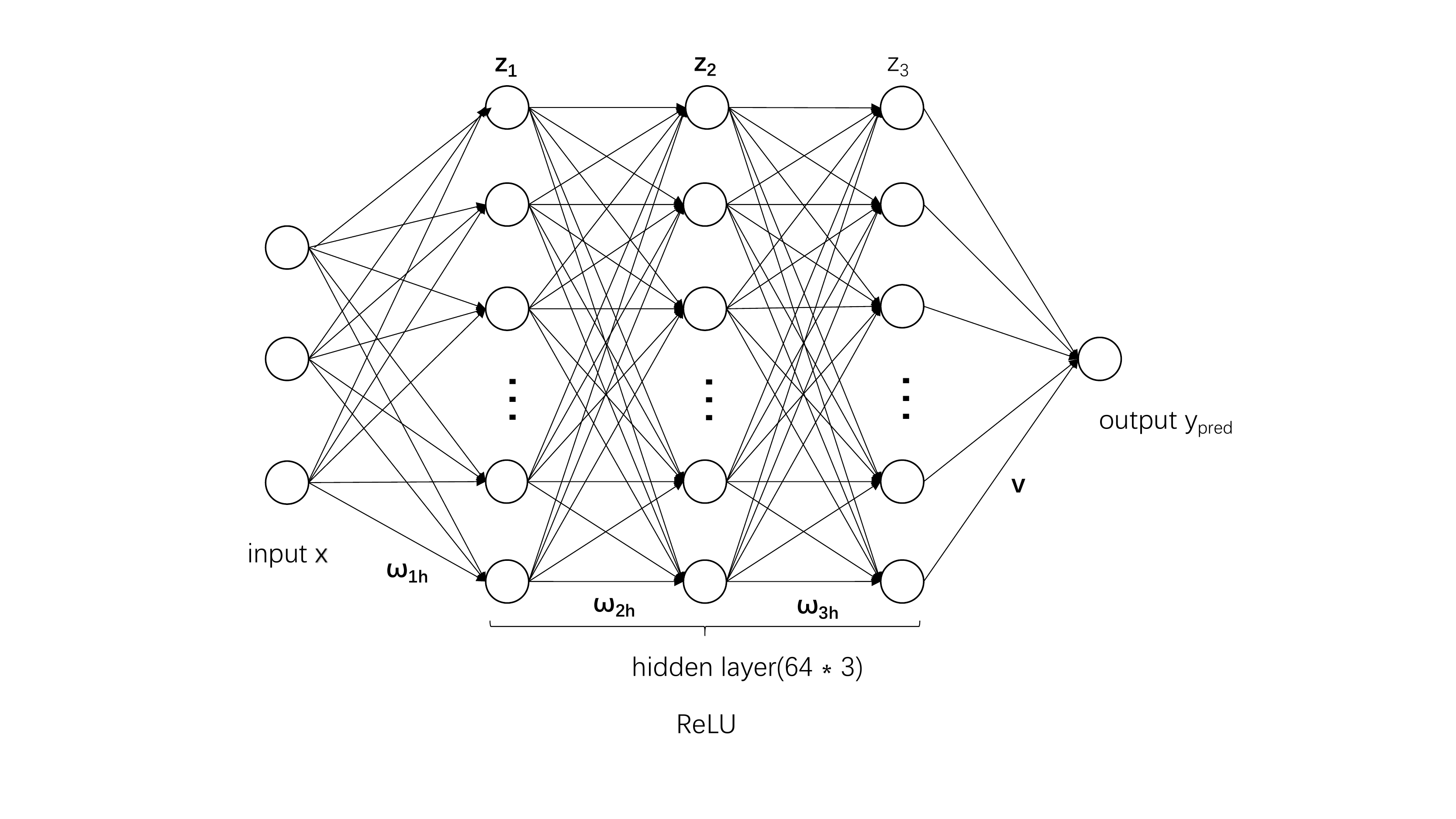}
\caption{Architecture of a typical fully connected feed-forward network having an input layer with certain units, three hidden layers each, containing 64 units, and a single output unit.}
\label{ANN}
\end{figure*}

As shown in Fig.~\ref{ANN}, we implement an MLP with three hidden layers for the $\alpha$-decay energy predictions. Rectified Linear Unit $(ReLU)$ function is chosen as the activation function with $ReLU(x)=log(1+exp(x))$. It is close to 0 when  $x_{i}$ is negative, and close to $x_{i}$ when $x_{i}$ is positive. In Fig.~\ref{ANN}, $\pmb{\omega}_{1h}$, $\pmb{\omega}_{2h}$ and $\pmb{\omega}_{3h}$ are the weight parameters belonging to the first, second and third hidden layers, respectively. The units of the first, second and third hidden layers are expressed as $\pmb{z_{1}}$, $\pmb{z_{2}}$, $\pmb{z_{3}}$ and \pmb{v} is the weight of the output layer. When input $x$ is entered into the input layer, the weighted sum is calculated, and the activation is propagated forward. The $ReLU$ function is selected as activation function,$z_{1h}$, is calculated as follows:
\begin{equation}
\begin{split}
z_{ih}&=ReLU(\pmb{\omega}_{ih}^{T} \pmb{x}) \\
      &=ReLU(\sum\limits_{j=1}^{d}\omega_{ihj} x_{j}),h=1,\ldots,H_{i},i=1,2,3 \\
\end{split}\label{2}
\end{equation}
where $H_{i}$ is the number of neurons, $\omega_{ihj}$ is the weight parameters in the hidden layer $i$ and d is the number of characteristic quantities in the input layer. When a pattern $\pmb{x}$ appears at the input, the system calculates a response based on two rules: first, the states of all neurons within a given layer, as specified by the outputs $z_{ih}$ of Eq.~(\ref{2}), are updated in parallel. Second, the layers are updated successively, precceding from the input to the output layer. Therefore, the output $y_{pred}$ is computed by taking $z_{3}$ as input. Thus, the forward propagation is completed. 
\begin{equation}
\begin{split}
y_{pred} = \textbf{v}^{T} \textbf{z}_{2}=\sum\limits_{h=1}^{H_{3}} v_{h}z_{3}+v_{0}
\end{split}
\end{equation}

In our ANN model, we use 64 hidden units in each hidden layer. The prediction for $\alpha$-decay energy is a single value and only one unit exists in the output layer. The challenge of machine learning demands learning model performs well not only in the training set, but also in the test set \citep{Heaton2017Ian}. As our results are shown below, the results obtained from ANN are in good agreement with the experimental data. The root-mean-square deviation between calculation with ANN model and the experimental value is very small. It is 0.09 MeV (0.135 MeV) for the training (test) data of ANN model with four inputs. It can be seen our neural network does not overfit, even though the number of its parameters is larger than the training data. 

The input data is randomly divided into two subsets as 80$\%$ for training and 20$\%$ for testing. The pragmatic objective of the training process will be to minimize the sum of squared errors $e_{t}$ relative to the experiment data. For the available experimental data $D=\left\lbrace (x_{1},y_{1}),(x_{2},y_{2}),\cdots,(x_{n},y_{n})\right\rbrace $, where $x_{i}$ and $y_{i} (i = 1, 2,..., n)$ are input and output data and $n$ is the number of data, the objective function is given as,

\begin{align}
E(\textbf{w},\textbf{v}|D)=\sum_{t=1}^{n}(e_{t})^2=\dfrac{\sum\limits_{t=1}^{n}(y_{pred}^{t}-y_{exp}^{t})^2}{n}
\end{align}
Here $y_{pred}$ is the output of ANN model, whereas $y_{exp}$ are experimental $\alpha$-decay energy.

We use Python.Keras to build ANN model and the Adam optimization algorithm is used to train our ANN model for 1000 epochs to minimize the mean square error. The name Adam comes from adaptive moment estimation. It is an adaptive gradient method, which adapts to the learning rate of model parameters alone. At the same time, we require a hyperparameter called Callsbacks.ReduceLROnPlateau in addition to learning rate \citep{htike2014unsupervised}. During training, we monitor the loss function. In the whole iteration process, the loss function is not reduced for 100 consecutive iterations, the callsback is activated. Then the gradient value in which loss function is minimum in the previous training process is reloaded. The reloaded gradient is then reduced by a factor of 0.1, and the model was continued to be optimized to find a smaller loss function. 
\begin{figure*}[htbp]
\includegraphics[scale=0.65]{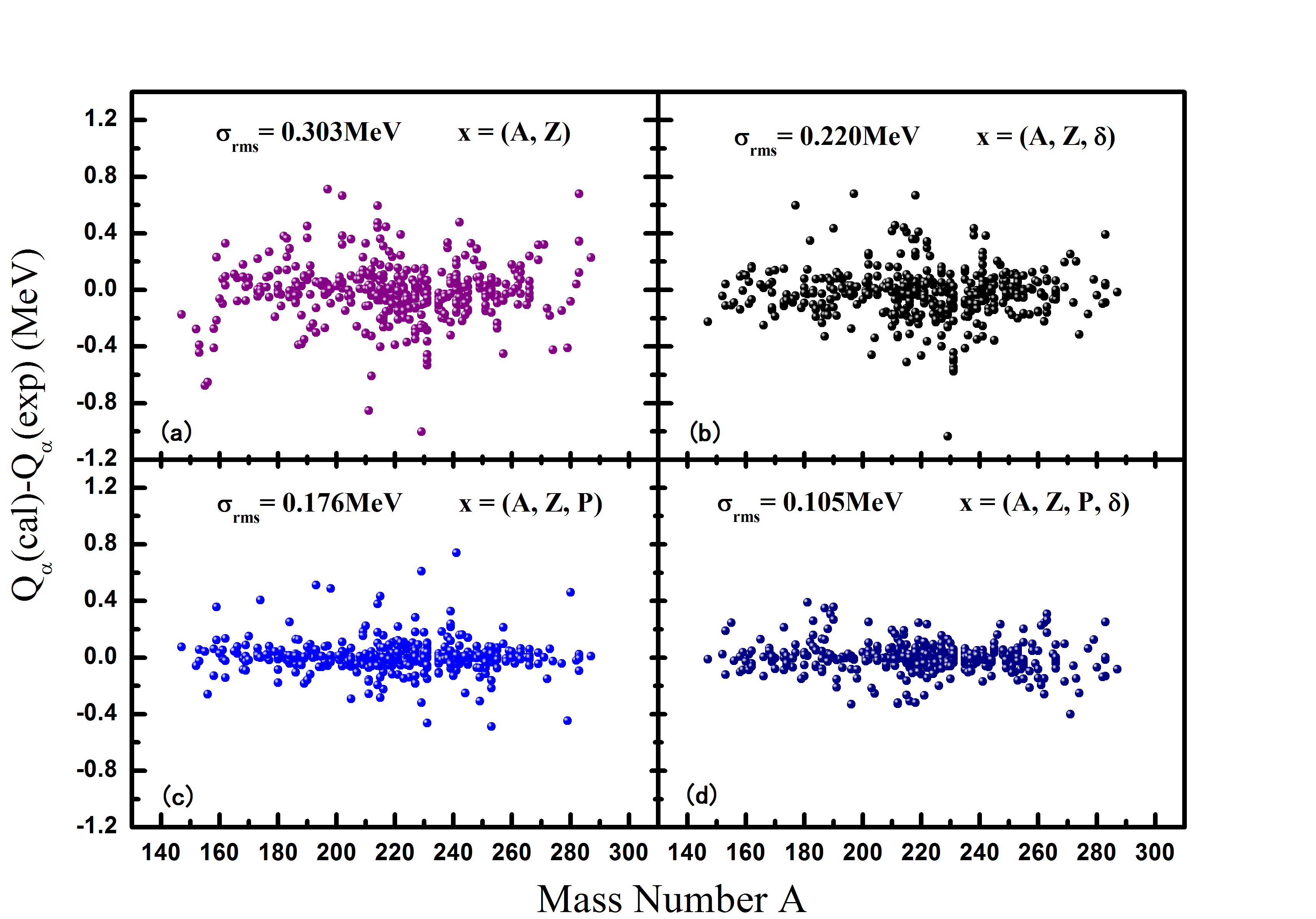}
\caption{(Color online) Comparison of the $Q_{\alpha}$ between the experimental data~\citep{timmurphy.org} and that calculated by the ANN model by using different inputs: (a) $\textbf{x}=(Z, A)$, (b) $\textbf{x}=(Z, A, \delta)$, (c) $\textbf{x}=(Z, A, P)$ and (d) $\textbf{x}=(Z, A, \delta, P)$. }
\label{alpha decay energy}
\end{figure*}

\section{Results and discussions}
\label{sec:Results}
\subsection{Prediction of the $\alpha$-decay energy based on the experimental data}

The experimental $\alpha$-decay energies are extracted from Te ($Z=52$) to Og ($Z=118$) isotopes. It includes not only the decay from the ground state of the parent nucleus to the ground state of the daughter nucleus, but also that to the exited state of the daughter nucleus. A total of 2131 $\alpha$-decay energy data are extracted, and it divided randomly into training set (80$\%$) and testing set (20$\%$).  
 
To improve the predictive power of ANN model, in addition to the mass (A) and proton (Z), more inputs which carry physical informations, should be included~\cite{E2021Nuclear}. Thus inputs $\delta$ and $P$ related to nuclear pairing and shell effects, respectively, are included. We consider these two effects separately and study their influences on the predictive performance of the ANN approach. The pairing $\delta(Z,N)$ is defined as 
\begin{equation}
\begin{split}
\delta(Z,N)=[(-1)^{Z}+(-1)^{N}]/2
\end{split}
\end{equation}
where $Z$ is the proton number and $N$ is the neutron number. A positive value of the pairing term indicates a more stable nucleus, while a negative value is the opposite\citep{Txt}. The shell effect $P$ \citep{2011Nuclear} reads,\begin{equation}
\begin{split}
P=\nu_{p}\nu_{n}/(\nu_{p}+\nu_{n}),
\end{split}
\end{equation}
where $\nu_{p(n)}$ is the difference between the atomic numbers and the closest magic number. We take the proton and neutron magic number as $Z=8,20,28,50,82,126$ and $N=8,20,28,50,82,126,184$, respectively. 

The standard deviations (in MeV) of calculated $Q_{\alpha}$ by ANN with respect to the available experiment values for different choices of input are shown in Fig. \ref{alpha decay energy}. In Figure. \ref{alpha decay energy} (a), we consider only mass A and proton Z as the inputs, the root-mean-square is slightly higher, reaching 0.303 MeV. In order to improve the predictive power of the ANN, we add more inputs by considering the physical effects which influence the $Q_{\alpha}$ strongly. By adding pairing effect $\delta$ alone, the root-mean-square deviation is significantly reduced and obtained as 0.22 MeV. It correspond to $50\%$ improvement in the prediction. By adding another input shell effect $P$ alone to the model, the root-mean-square deviation is 0.170 MeV. Compared with pairing effect $\delta$, including shell effect $P$ improves better of prediction accuracy of the ANN model. When we consider both the pair effect $\delta$ and the shell effect $P$, the root-mean-square reaches a minimum value of 0.105 MeV. The results indicate that when we take appropriate physical features as inputs in the ANN model, one can get more accurate predictions of $\alpha$-decay energy.

\renewcommand{\arraystretch}{1.5}
\setlength{\tabcolsep}{5pt}
\LTcapwidth=170mm
\begin{longtable*}{cccccccc}
\caption{The root-mean-square deviation($\sigma_{rms}$) in unit of MeV(modified the table as we discussed several days ago).}
\label{tab}\\
\hline
                         &               & \multicolumn{4}{c}{ANN model}                 & XGBoost\citep{SaxenaG2021_JPGNPP48_55103}      & DZ+BNN model \citep{2019Alpha}       \\ \hline
\endfirsthead
\multicolumn{8}{c}%
{{\bfseries Table \thetable\ continued from previous page}} \\
\hline
                           &               & \multicolumn{4}{c}{ANN model}                 & DZ model       & BNN       \\ \hline
\endhead
input \textbf{x}                            &               & ($A,Z$) & ($A,Z$,$\delta$) & ($A,Z,P$) &($A,Z,P$,$\delta$) & - & ($A,Z,P$) \\ \hline
\multirow{2}{*}{$\sigma_{rms}$(MeV)} & training set & 0.150    & 0.135     & 0.115     & 0.090        & -          & 0.178     \\
                           & test set     & 0.303   & 0.220      & 0.176     & 0.105       & 0.403          & 0.274     \\ \hline
\end{longtable*}

The root-mean-square deviations ($\sigma_{rms}$) of the $\alpha$-decay energy by different machine learning methods are given in Table. \ref{tab}. The present calculation exhibits high predictive accuracy of ANN model. For test set, except for $\sigma_{rms}=0.303$ which is obtained by input $\textbf{x}=(A,Z)$, all the other $\delta_{rms}$ deviations are smaller than that given by the DZ+BNN model ($\sigma_{rms}=0.274$)~\cite{2019Alpha} and XGBoost neural network ($\sigma_{rms}=0.310$)~\cite{SaxenaG2021_JPGNPP48_55103}. When input $\textbf{x}=(A,Z,P)$ is adopted, which is as same as that used by DZ+BNN model, $\sigma_{rms}=0.176$ is obtained in the ANN calculations. In the present investigation, we extract not only the $\alpha$-decay energy $Q_{\alpha}$ from the ground state of the parent nucleus to the ground state of the daughter, but also to the excited state of the daughter nucleus. It increases the number of sample data and improve predictive power of the ANN model. When further considering the pairing effect $\delta$, whereas BNN model do not take into account this effect~\citep{2019Alpha}, the results of ANN model is improved to $\sigma_{rms}=0.105$. This shows a very high predictive power. Since the most powerful prediction is given by using input $\textbf{x}=(Z, A, \delta, P)$, the following calculations used to discuss the $\alpha$-decay energy in the SHE region are all calculated by using input $\textbf{x}=(Z, A, \delta, P)$.  

\vspace{10pt}

\subsection{Extrapolation of the $\alpha$-decay energy in the superheavy nuclei mass region}
\vspace{-3pt}

In the superheavy nuclei mass region where there is no sufficient experimental data available, accurate prediction of the $\alpha$-decay energy and the half lives of $\alpha$-decay is very important both for the the synthesis of the new superheavy elements and the structural study of superheavy nuclei. From the above discussion, one can seen that the ANN model have a comparatively high predictive power of $\alpha$-decay energy. We apply the above ANN approach to calculate the $\alpha$-decay energy of SHE regions. The results are listed in Table. \ref{tab2}. The root-mean-square of all the nuclei calculated in SHE region is 0.204 MeV. The root-mean-square of each element is below 0.320 MeV, which is comparable to the results obtained by the theoretical studies in Refs.~\citep{PhysRevC.100.061302,PhysRevC.85.054303}. The element with the largest root-mean-square deviation is Mt isotopes ($\sigma_{rms}= 0.310$ MeV). Its effect on the half-life is about 1 to 2 orders of magnitude. It confirms us that the ANN model can give good prediction of the $Q_{\alpha}$ values in the SHE region. 

\renewcommand{\arraystretch}{1.5}
\setlength{\tabcolsep}{3.5pt}
\LTcapwidth=170mm
\begin{longtable*}{@{}ccccccccccccccc@{}}
\caption{The root-mean-square deviations (in MeV) of the $\alpha$-decay energy between the calculation with ANN model and the experimental data for the SHE mass region with $104 \leqslant Z \leqslant 118$. }
\label{tab2}\\\hline
\toprule
element & number & $\sigma_{rms}$ (MeV) &  & element & number & $\sigma_{rms}$ (MeV) &  & element & number & $\sigma_{rms}$ (MeV) &  & element & number & rms (MeV) \\* \midrule
\endhead
\bottomrule
\endfoot\hline
\endlastfoot\hline
all & 87 & 0.204  &  & Rf & 5 & 0.167 &  & Db & 9 & 0.127 &  & Sg & 7 & 0.177  \\
Bh  & 8  & 0.179  &  & Hs & 9 & 0.235 &  & Mt & 7 & 0.310  &  & Ds & 8 & 0.289  \\
Rg  & 7  & 0.271  &  & Cn & 4 & 0.182 &  & Nh & 6 & 0.132 &  & Fr & 5 & 0.141  \\
Mc  & 4  & 0.092 &  & Lv & 6 & 0.106 &  & Ts & 3 & 0.091 &  & Og & 1 & 0.164 \\* \bottomrule
\end{longtable*}

\begin{figure*}[htbp]
\includegraphics[scale=0.65]{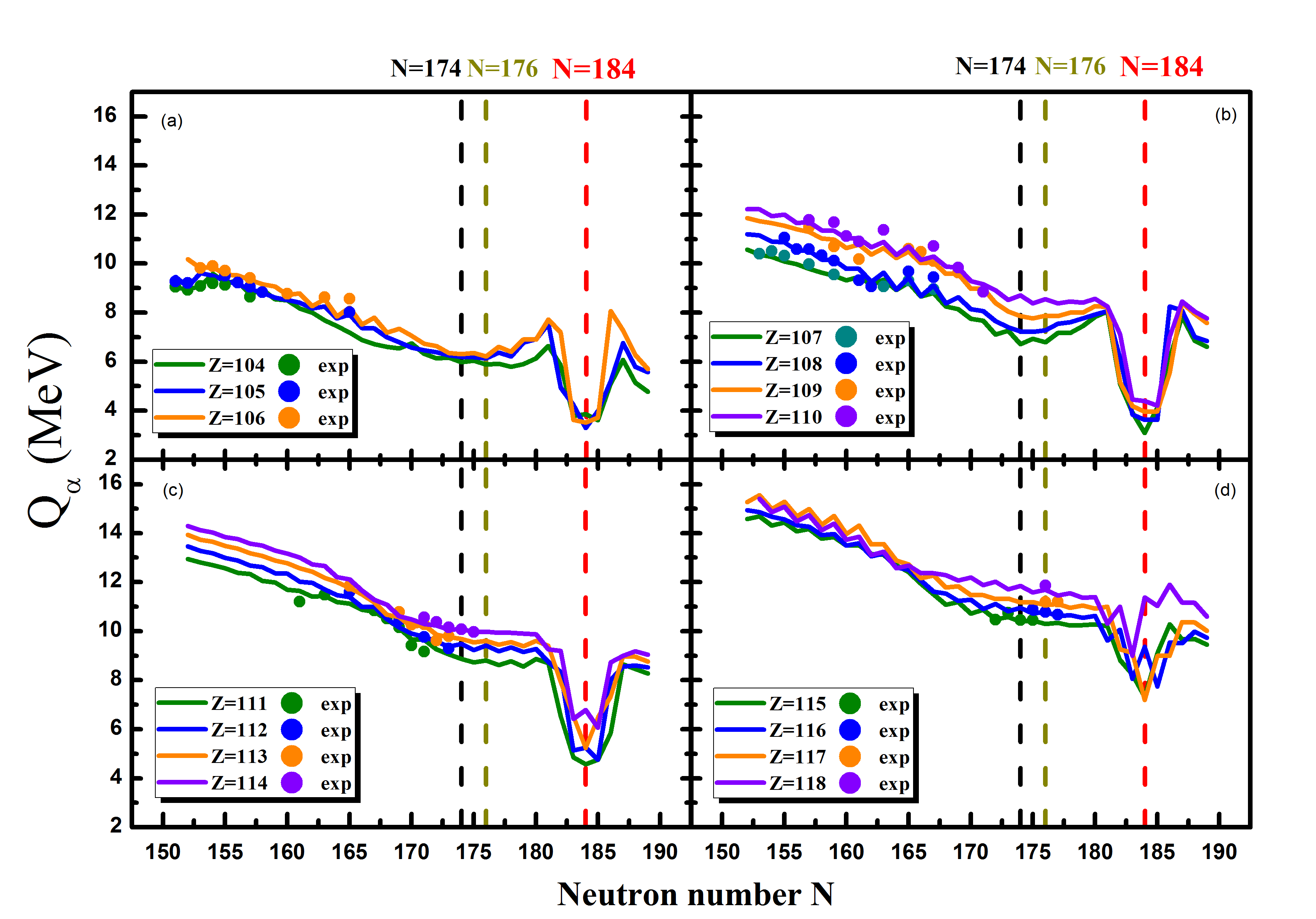}
\caption{(Color online) The $\alpha$-decay energy $Q_{\alpha}$ as a function of neutron numbers $N$ in the SHE mass region with $104 \leqslant Z \leqslant118$. The results obtained by the ANN model are denoted by lines and the experimental data are by solid circle. The dashed vertical lines are used to mark the predicted possible neutron shell gaps.}\label{shn}
\end{figure*}

The detailed comparison of the calculated $Q_{\alpha}$ with the available experimental data for $Z=104-118$ isotope chains are shown in Fig.~\ref{shn}, in which the results are divided into four groups. The neutron numbers are vary from $N=151$ to $189$. The ANN calculated results are denoted by the solid lines and the experimental data by solid circles. One can see that the experimental data are reproduced well by the ANN model. The local minimum of the $Q_{\alpha}$ curves with the neutron number $N$ of the parent nucleus could indicate a magic or sub-magic number. The dashed vertical lines mark the neutron numbers, at which there are possible existent shell gaps. 

As shown in Fig. \ref{shn} that obvious minimum appears at $N = 184$ for almost all the $Z=104-118$ isotope chains, which indicates a big shell gap. In Fig. \ref{shn} (a), $Q_{\alpha}$ decreases gradually from $N=151$ to around 174 or 176, then increased until $N=180$. The local minimum indicates a sub-shell gap at $N=174$ or 176. In Fig. \ref{shn} (b), although the minimum at $N=174(176)$ is not as obvious as that for $Z=104-106$, it still appears to exist. That means a possible sub-shell gap at $N=174(176)$ for $Z=107-110$ isotopes. As for the $Z=111-114$ and $Z=115-118$ isotopes shown in Fig.~\ref{shn} (c) and (d), respectively, $Q_{\alpha}$ decreases from $N=151$ to 174 fast and then becomes flat as neutron number increased to $N=180$. There is no sub-shell gap shown at $N=174(176)$. For $Z = 118$, the minimum of $Q_{\alpha}$ is predicted at $N$ = 183 instead of 184. This is partly because there is little experimental data for Og isotopes in the training model. For the magic number at $N=184$ and the sub-shell gap at $N=174(176)$, further experimental and theoretical investigations are needed. 


\section{Summary}
\label{sec:Summary}

In the present work, we build and train the ANN model by extracting experimental $Q_{\alpha}$ from the ground state of parent nuclei to the ground state of daughter nuclei and to the excited state. By this way, the number of samples is increased substantially. The ANN model would be trained to have more prediction power. To obtain a high predictive power, besides mass number $A$ and proton number $Z$, two more inputs, i.e., P and $\delta$ which is related to nuclear shell effect and pair effect, respectively, are introduced. By studying the 2131 $\alpha$-decays, the root-mean-square of the $\alpha$-decay energy is 0.105, which presents a great accuracy. The influence of different input on the predictive power is investigated. It is found that either the shell effect or the pairing effect could leads to an obvious improvement of the result, and the shell effect plays a more important role. The optimal result is obtained as both the shell and pairing effects are considered simultaneously. The ANN model is used to study the $\alpha$-decay energies in superheavy nuclear mass region where experimental data are rare. The ANN results can reproduce the available experimental data very well. The predicted $Q_{\alpha}$ for SHE region suggests the neutron magic number at $N=184$ and the possible sub-shell gaps around $N=174(176)$. 

\begin{acknowledgements}
We thank Prof. Zhong-Ming Niu in Anhui University for the helpful discussion. This work is supported by the National Natural Science Foundation of China (Grant Nos. U2032138, 11775112) and the National College Students Innovation and Entrepreneurship Training Program (Grant No. 202110287042).
\end{acknowledgements}

\bibliographystyle{apsrev4-1}
\bibliography{references}

\end{document}